\documentclass[aps,prl,twocolumn,superscriptaddress,amsmath,amssymb,floatfix,longbibliography]{revtex4-2}

\usepackage{graphicx}
\usepackage{bm}
\usepackage{hyperref}
\usepackage{xcolor}
\colorlet{BLUE}{blue}
\usepackage{tikz}
\usetikzlibrary{
	decorations.pathmorphing,
	decorations.pathreplacing,
	decorations.markings,
	arrows.meta
}

\hypersetup{
	colorlinks=true,
	linkcolor=blue,
	filecolor=magenta,      
	urlcolor=blue,
	citecolor=blue,
}

\begin{document}
	
	\title{Geometric Rashba Control of Polar Pairing at a-LaAlO$_3$/KTaO$_3$ Interfaces}
	
	\author{Yi Zhou}
	\email{yizhou@iphy.ac.cn}
	\affiliation{Institute of Physics, Chinese Academy of Sciences, Beijing 100190, China}
	
	\date{\today}
	
	\begin{abstract}
		At amorphous-LaAlO$_3$/KTaO$_3$ interfaces, superconductivity coexists with switchable polar nanoregions (PNRs), and $T_c$ depends quasi-linearly on crystallographic orientation. We propose a minimal theoretical framework in which overdamped PNR fluctuations mediate pairing, while geometric Rashba coupling controls its angular dependence. Within a reduced isotropic helicity-band description, the leading dynamic Rashba vertex scales as $\sin\theta$, where $\theta$ is the angle between the interface normal and the $[100]$ axis, yielding $\lambda(\theta)=\lambda_0+C\sin^2\theta$. Numerical Matsubara--Eliashberg solutions show that this nonlinear mapping reproduces the observed quasi-linear $T_c(\theta)$. Because the Rashba-activated polar channel is amplified by the large atomic spin-orbit coupling of Ta $5d$ orbitals, the framework also offers a parametric---not yet quantitative---rationale for why KTaO$_3$ interfaces are both more orientation-sensitive and higher-$T_c$ than their SrTiO$_3$ counterparts.
	\end{abstract}
	
	\maketitle
	
	\textbf{Introduction.---} Two-dimensional (2D) superconductivity at KTaO$_3$ (KTO)-based oxide interfaces, including LaAlO$_3$/KTaO$_3$ (LAO/KTO), is strikingly sensitive to crystallographic geometry: the transition temperature $T_c$ varies strongly with interface orientation, peaking on high-index facets and collapsing at the pristine (100) surface \cite{Liu2021, Chen2021Science, Chen2021PRL, Cao2026}. KTO interfaces also exhibit substantially higher $T_c$ than their SrTiO$_3$ (STO) counterparts, reaching $T_c\sim2$ K on KTO (111) \cite{Cao2026, Cheng2025}, compared with $T_c\sim0.2$--$0.3$ K in LAO/STO \cite{Reyren2007, Caviglia2008}. Since polar modes have been proposed as an important pairing channel at STO interfaces \cite{Enderlein2020}, it is natural to ask whether a related channel contributes to the higher and more geometry-dependent $T_c$ of KTO interfaces.
	
	Two experimental observations sharpen this puzzle. First, across the high-index amorphous-LaAlO$_3$/KTaO$_3$ (a-LAO/KTO) facets analyzed here, $T_c$ varies quasi-linearly with an orientation angle $\theta$ \cite{Cao2026}; related orientation dependence has also been observed in ionic-liquid-gated KTO interfaces \cite{Ren2022}. For a symmetry-equivalent representative facet $(hkl)$ chosen with $h\ge k\ge l\ge0$, $\theta=\arccos[h/\sqrt{h^2+k^2+l^2}]$ denotes the angle between the interface normal and $[100]$, ranging from $0^\circ$ at (100) to $\arccos(1/\sqrt{3})\approx54.7^\circ$ at (111) \cite{SupplementalMaterial}. Second, transport hysteresis and Raman scattering indicate collective ordering of polar nanoregions (PNRs) below $\sim50$ K \cite{Zhang2026}, while piezoresponse-force microscopy provides evidence that switchable out-of-plane interfacial polarity couples to superconductivity \cite{Dong2026}. Together, these observations motivate a pairing scenario involving a soft interfacial polar mode whose coupling to electrons is controlled by interface geometry in the presence of broken inversion symmetry.
		
	An electronic-dipole-fluctuation theory \cite{Palle2024} has highlighted the roles of parity mixing and spin-orbit coupling (SOC), but its weak-coupling treatment does not by itself address either the enhanced $T_c$ scale of KTO or the observed quasi-linear $T_c(\theta)$. Motivated by these observations, we construct a minimal theoretical framework in which overdamped interfacial PNR fluctuations mediate pairing and a geometrically controlled dynamic Rashba vertex produces the angular dependence. Within the reduced model, the slow PNR dynamics provide broad low-energy spectral weight, the leading-order vertex scales as $\sin\theta$, and projection onto one helicity band yields an effective scalar intra-band kernel. The resulting SOC-amplified polar channel reproduces the observed $T_c(\theta)$ trend and offers a parametric, rather than quantitative, rationale for the larger $T_c$ scale of KTO relative to STO.

	\textbf{Nature of the Soft Bosonic Mode.---} Bulk KTO is a quantum paraelectric with a soft, strongly anharmonic transverse-optical (TO) mode whose energy saturates near $3$ meV at low temperature \cite{Shirane1967}. Quantum fluctuations prevent long-range polar order, although local polar correlations persist \cite{Ichikawa2005, Aktas2014}. The bulk is insulating; by contrast, the a-LAO/KTO interface combines these polar degrees of freedom with a 2D electron gas of Fermi energy $E_F$ under a built-in inversion-breaking field. Experimentally, atomic-resolution scanning transmission electron microscopy reveals an out-of-plane distortion \cite{Dong2026}, while Raman modes and bistable transport below $\sim50$ K indicate collective polar ordering \cite{Zhang2026}; together, these observations provide evidence that interfacial polarity coexists with superconductivity. Motivated by this evidence, we assume that the confining field preferentially aligns the interfacial PNR dipoles out of plane.
	
	We take local finite-$\bm{q}$ amplitude fluctuations $\delta P_z$ of these out-of-plane PNR dipoles to be the bosonic modes mediating Cooper pairing (Fig.~\ref{fig:schematic}). Depolarizing fields stiffen the uniform ($\bm{q}\to0$) out-of-plane polarization mode of the interfacial layer \cite{Cochran1960, Junquera2003}; we assume, however, that nonuniform finite-$\bm{q}$ fluctuations remain soft and dynamically modulate the confining potential.
	
	\begin{figure}[tb]
		\centering
		\begin{tikzpicture}[
			x=0.70cm,
			y=0.70cm,
			line cap=round,
			line join=round,
			font=\sffamily\scriptsize,
			>=Latex
		]
			\definecolor{laatom}{HTML}{11AFC0}
			\definecolor{alatom}{HTML}{E7BC16}
			\definecolor{katom}{HTML}{3B9E55}
			\definecolor{taatom}{HTML}{3558A8}
			\definecolor{oatom}{HTML}{C94235}
			\definecolor{polar}{HTML}{D97706}
			\definecolor{bulk}{HTML}{2F7D4C}

			\path[use as bounding box] (0,-1.30) rectangle (11.70,8.72);

			\node[anchor=west,font=\sffamily\bfseries\footnotesize]
				at (0.12,8.52) {(a) Interface structure};

			\foreach \x/\atomcolor/\atomname in {
				4.85/laatom/La,
				6.05/alatom/Al,
				7.25/katom/K,
				8.35/taatom/Ta,
				9.55/oatom/O
			}{
				\fill[\atomcolor] (\x,8.52) circle (.11);
				\draw[\atomcolor!70!black,line width=.30pt]
					(\x,8.52) circle (.11);
				\node[anchor=west,font=\sffamily\scriptsize]
					at (\x+.18,8.52) {\atomname};
			}

			\fill[laatom!5,rounded corners=2pt] (2.72,7.25) rectangle (11.15,8.28);
			\fill[bulk!4] (2.72,2.95) rectangle (11.15,7.15);
			\fill[polar!10] (2.72,4.35) rectangle (11.15,7.15);
			\draw[polar!70,densely dashed,line width=.55pt]
				(2.72,4.35) -- (11.15,4.35);
			\draw[black,line width=1pt] (3.02,7.15) -- (11.15,7.15);

			\node[anchor=west,font=\sffamily\bfseries\scriptsize]
				at (0.18,7.91) {a-LAO};
			\node[anchor=west,font=\sffamily\scriptsize,text=gray!75!black]
				at (0.18,7.63) {amorphous};

			\node[anchor=west,font=\sffamily\bfseries\scriptsize,text=polar!85!black]
				at (0.18,6.82) {2D electron gas};

			\node[anchor=west,font=\sffamily\bfseries\scriptsize]
				at (0.18,5.75) {polar KTO};
			\node[anchor=west,font=\sffamily\scriptsize,text=gray!75!black]
				at (0.18,5.45) {aligned PNRs};

			\node[anchor=west,font=\sffamily\bfseries\scriptsize]
				at (0.18,3.48) {bulk KTO};

			\draw[polar!85!black,line width=.55pt]
				(11.35,4.35) -- (11.35,7.15);
			\draw[polar!85!black,line width=.55pt]
				(11.23,4.35) -- (11.47,4.35)
				(11.23,7.15) -- (11.47,7.15);
			\node[rotate=90,font=\sffamily\scriptsize,text=polar!85!black]
				at (11.50,5.75) {$d_{\mathrm{sc}}$};

			\draw[->,gray!75!black,line width=.55pt]
				(11.35,3.05) -- (11.35,3.45);
			\node[anchor=south,font=\sffamily\tiny,text=gray!75!black]
				at (11.35,3.47) {$+\hat z$};

			\foreach \x/\y/\r in {
				3.00/7.51/.13,3.64/7.80/.16,4.30/7.50/.14,4.96/7.92/.16,
				5.62/7.57/.15,6.28/7.87/.16,6.94/7.50/.14,7.60/7.82/.16,
				8.26/7.56/.15,8.92/7.91/.16,9.58/7.51/.14,10.24/7.82/.16
			}{
				\fill[laatom] (\x,\y) circle (\r);
				\fill[white,opacity=.42] (\x-.04,\y+.05) circle (.03);
			}
			\foreach \x/\y in {
				3.24/7.97,3.94/7.55,4.62/7.78,5.30/7.43,5.98/8.03,
				6.66/7.67,7.34/8.00,8.02/7.43,8.70/7.74,9.38/7.44,
				10.06/8.00,10.74/7.60
			}{
				\fill[alatom] (\x,\y) circle (.09);
			}
			\foreach \x/\y in {
				2.84/7.88,3.40/7.40,3.74/8.10,4.08/7.94,4.42/7.39,
				4.76/8.12,5.10/7.70,5.44/8.09,5.78/7.42,6.12/7.58,
				6.46/8.14,6.80/7.86,7.14/7.39,7.48/7.62,7.82/8.10,
				8.16/7.91,8.50/7.40,8.84/8.12,9.18/7.66,9.52/8.13,
				9.86/7.82,10.20/7.39,10.54/8.13,10.88/7.93
			}{
				\fill[oatom] (\x,\y) circle (.065);
			}

			\foreach \y/\shift in {3.65/0,5.05/.06,6.45/.12}{
				\foreach \x in {3.55,5.15,6.75,8.35,9.95}{
					\coordinate (octTop) at (\x,\y+.74);
					\coordinate (octBottom) at (\x,\y-.74);
					\coordinate (octWest) at (\x-.83,\y+.02);
					\coordinate (octEast) at (\x+.83,\y-.02);
					\coordinate (octRear) at (\x+.28,\y+.22);
					\coordinate (octFront) at (\x-.28,\y-.22);
					\coordinate (octTa) at (\x,\y+\shift);

					\fill[oatom!5] (octTop) -- (octRear) -- (octWest) -- cycle;
					\fill[oatom!12] (octTop) -- (octEast) -- (octRear) -- cycle;
					\fill[oatom!10] (octBottom) -- (octWest) -- (octRear) -- cycle;
					\fill[oatom!16] (octBottom) -- (octRear) -- (octEast) -- cycle;
					\fill[oatom!25] (octTop) -- (octWest) -- (octFront) -- cycle;
					\fill[oatom!38] (octTop) -- (octFront) -- (octEast) -- cycle;
					\fill[oatom!31] (octBottom) -- (octFront) -- (octWest) -- cycle;
					\fill[oatom!45] (octBottom) -- (octEast) -- (octFront) -- cycle;

					\draw[oatom!74!black,line width=.47pt]
						(octTop) -- (octWest) -- (octBottom) -- (octEast) -- cycle;
					\draw[oatom!67!black,line width=.39pt]
						(octTop) -- (octFront) -- (octBottom)
						(octWest) -- (octFront) -- (octEast);
					\draw[oatom!52,densely dashed,line width=.31pt]
						(octTop) -- (octRear) -- (octBottom)
						(octWest) -- (octRear) -- (octEast);

					\draw[taatom!75!black,line width=.27pt,opacity=.82]
						(octTa) -- (octTop)
						(octTa) -- (octBottom)
						(octTa) -- (octWest)
						(octTa) -- (octEast)
						(octTa) -- (octFront);
					\draw[taatom!68!black,densely dashed,line width=.25pt,opacity=.72]
						(octTa) -- (octRear);

					\shade[ball color=oatom!82!black] (octRear) circle (.058);
					\shade[ball color=oatom] (octTop) circle (.078);
					\shade[ball color=oatom] (octBottom) circle (.078);
					\shade[ball color=oatom] (octWest) circle (.078);
					\shade[ball color=oatom] (octEast) circle (.078);
					\shade[ball color=oatom] (octFront) circle (.078);
					\shade[ball color=taatom] (octTa) circle (.120);
				}
			}

			\foreach \x in {4.35,5.95,7.55,9.15}{
				\shade[ball color=katom] (\x,5.75) circle (.125);
			}

			\draw[->,polar,line width=.9pt]
				(10.92,7.78) -- (10.92,6.74)
				node[pos=.40,right=1pt,text=black] {$E_0$};

			\begin{scope}[shift={(0,-1.50)}]
				\node[anchor=west,font=\sffamily\bfseries\footnotesize]
					at (0.12,4.02) {(b) Local polar dynamics};

				\fill[polar!5,rounded corners=2pt] (0.12,0.28) rectangle (5.55,3.76);
				\draw[polar!75!black,rounded corners=2pt,line width=.65pt]
					(0.12,0.28) rectangle (5.55,3.76);
				\fill[polar] (0.12,3.62) rectangle (5.55,3.76);
				\node[font=\sffamily\bfseries\footnotesize]
					at (2.84,3.34) {interfacial PNR};
				\node[font=\sffamily\scriptsize,text=polar!85!black]
					at (2.84,2.91) {overdamped asymmetric mode};

				\draw[->,gray!70] (0.65,.77) -- (5.05,.77);
				\node[anchor=north east,font=\sffamily\tiny,text=gray!75!black]
					at (5.35,.72) {$P_z$};
				\draw[->,gray!70] (0.82,.61) -- (0.82,2.62);
				\node[anchor=east,font=\sffamily\tiny,text=gray!75!black]
					at (0.68,2.30) {$U$};
				\draw[black,line width=.85pt,samples=80,domain=1.18:4.78]
					plot (\x,{1.10 + .22*((\x-3.00)^2-.95)^2
						+ .075*(\x-3.00)});
				\fill[taatom] (2.04,1.00) circle (.09);
				\draw[<->,polar!90!black,line width=.55pt,
					>={Latex[length=1.8pt,width=2.2pt]},
					preaction={draw=polar!5,line width=2.2pt}]
					(1.83,.77) -- (2.25,.77);
				\node[font=\sffamily\tiny,text=polar!90!black]
					at (2.04,.49) {$\delta P_z$};

				\fill[bulk!4,rounded corners=2pt] (5.80,0.28) rectangle (11.23,3.76);
				\draw[bulk!80!black,rounded corners=2pt,line width=.65pt]
					(5.80,0.28) rectangle (11.23,3.76);
				\fill[bulk] (5.80,3.62) rectangle (11.23,3.76);
				\node[font=\sffamily\bfseries\footnotesize]
					at (8.52,3.34) {bulk KTO TO phonon};
				\node[font=\sffamily\scriptsize,text=bulk!90!black,align=center]
					at (8.52,2.72) {soft anharmonic mode\\(quantum paraelectric)};

				\draw[->,gray!70] (6.33,.77) -- (10.73,.77);
				\node[anchor=north east,font=\sffamily\tiny,text=gray!75!black]
					at (11.03,.72) {$P_z$};
				\draw[->,gray!70] (6.50,.61) -- (6.50,2.62);
				\node[anchor=east,font=\sffamily\tiny,text=gray!75!black]
					at (6.36,2.30) {$U$};
				\draw[black,line width=.85pt,samples=120,domain=6.90:10.52]
					plot (\x,{1.02 + .105*((\x-8.71)^2)^2 - .035*(\x-8.71)^2});
				\fill[taatom] (8.71,1.04) circle (.09);
				\draw[<->,bulk!90!black,line width=.55pt,
					>={Latex[length=1.8pt,width=2.2pt]},
					preaction={draw=bulk!4,line width=2.2pt}]
					(8.28,.77) -- (9.14,.77);
				\node[font=\sffamily\tiny,text=bulk!90!black]
					at (8.71,.49) {$\delta P_z$};
			\end{scope}
		\end{tikzpicture}
		\caption{Schematic of the a-LAO/KTO interface and the modeled polar dynamics. (a) The 2D electron gas occupies the superconducting interfacial layer of thickness $d_{\mathrm{sc}}$ on the KTO side. The model assumes that the built-in field $E_0$ preferentially aligns the interfacial PNR dipoles out of plane; deeper in bulk KTO, Ta ions remain centered on average. (b) Schematic potential profiles (not calculated energy surfaces) represent the interfacial amplitude fluctuation $\delta P_z$ as an overdamped mode in an asymmetric double well and the bulk soft TO mode in a shallow, strongly anharmonic potential where quantum fluctuations prevent long-range polar order.}
		\label{fig:schematic}
	\end{figure}
	
	Carrier-induced Landau damping and disorder from the amorphous LAO overlayer provide plausible sources of damping for these soft finite-$\bm q$ polar fluctuations. We model a soft finite-$\bm q$ sector of their response using a causal, overdamped second-order Brownian oscillator:
	\begin{equation} \label{eq:chi}
		\chi(\bm{q}, \omega) = \frac{\chi_0 \Omega_B^2}{\Omega_B^2(1 + \xi_P^2 \bm{q}^2) - \omega^2 - i \gamma \omega},
	\end{equation}
	Here $\chi_0$ sets the static susceptibility scale, $\xi_P$ is an effective PNR correlation length controlling spatial dispersion, $\Omega_B$ is the oscillator frequency parameter, and $\gamma$ is the damping rate. For the isotropic calculation below, we approximate the pairing-weighted momentum average of Eq.~\eqref{eq:chi} by a single normalized Brownian spectrum, with unresolved momentum dependence absorbed into its effective parameters. Within the near-threshold Allen--Dynes family discussed in the Supplemental Material, changes in $\Omega_{\log}$, $\lambda_0$, and $C$ compensate one another, yielding nearly indistinguishable $T_c(\theta)$ curves; we adopt $\Omega_{\log}=0.84$ meV as a representative scale \cite{SupplementalMaterial}. Taking $\Omega_B=3$ meV, motivated by the bulk TO scale, then determines $\gamma\approx8.8$ meV. The resulting effective spectrum is overdamped, has a Debye-like low-energy response, and satisfies $\operatorname{Im}\chi\sim\omega^{-3}$ at high frequencies, ensuring a finite second Eliashberg moment.
	
	\textbf{Geometric Tuning and Helicity Projection.---} The coupling between polar fluctuations and conduction electrons is constrained by inversion and orbital symmetry. Bulk KTO is centrosymmetric, whereas the interfacial confining potential breaks inversion symmetry and permits a dynamic Rashba vertex. A microscopic treatment of an arbitrary $(hkl)$ interface would require anisotropic hopping, orbital mixing, and confinement-induced subband reordering; here we use a reduced orbital-rotation construction to isolate the leading angular dependence.
	
	Within this reduced construction, the normal-field dipole matrix element of a representative $|d_{yz}\rangle$-derived orbital channel vanishes at (100) by parity. Tilting the interface normal away from $[100]$ by $\theta$ mixes in the normal-field-active orbital component with amplitude $\sin\theta$, giving the leading dependence \cite{SupplementalMaterial}:
	\begin{equation} \label{eq:g_theta}
		g(\theta)=g_0\sin\theta.
	\end{equation}
	Here $g_0$ is treated as orientation independent within the minimal model. Other orbitals, subband reordering, and facet-dependent confinement can renormalize its coefficient or generate additional angular terms.
	
	The normal component $P_{\hat n}(\bm q)$ of the polar fluctuation, with $\hat n$ the unit interface normal, is denoted $\delta P_z$ in Fig.~\ref{fig:schematic} and dynamically modulates the inversion-breaking confining field and hence the Rashba coupling. This gives a Kozii--Fu-type electron-boson interaction \cite{Kozii2015}:
	\begin{equation} \label{eq:KoziiFu}
		H_{\text{int}} = g(\theta) \sum_{\bm{k}, \bm{q}} P_{\hat{n}}(\bm{q})
		\psi^\dagger_{\bm{k}+\bm{q}/2}
		\left[ (\bm{k} \times \bm{\sigma}) \cdot \hat{n} \right]
		\psi_{\bm{k}-\bm{q}/2}.
	\end{equation}
	Here $\psi_{\bm k}$ is the electron spinor and $\bm{\sigma}$ denotes the Pauli matrices. The resulting orientation dependence is consistent with electron-phonon matrix elements observed by angle-resolved photoemission spectroscopy (ARPES) \cite{Chen2024NatCommun}.
	
	Because the interaction has the spin structure of the static Rashba term, its projection onto Rashba helicity eigenstates is predominantly intra-band at small $\bm q$. We therefore approximate the momentum-dependent spinor overlap within the lowest helicity band by an isotropic Fermi-surface average, obtaining $\Gamma_{\mathrm{eff}}\approx-g(\theta)k_F$, where $k_F$ is a representative Fermi wave vector. This scalar reduction parameterizes the integrated pairing scale but does not determine the mixed-parity gap structure \cite{SupplementalMaterial}.
	
	\textbf{Migdal-Eliashberg Formulation.---} Integrating out the polar fluctuations gives the retarded effective interaction shown diagrammatically in Fig.~\ref{fig:Theory}(a):
	\begin{equation}
		V_{\text{eff}}(\bm{k}, \bm{k}', \omega) = |\Gamma_{\text{eff}}|^2 \chi(\bm{k}-\bm{k}', \omega).
	\end{equation}
		
	\begin{figure}[tb]
		\centering
		\begin{tabular}{@{}c@{\hspace{0.2cm}}c@{}} 
			\textbf{(a)} & \textbf{(b)} \\
			\begin{tikzpicture}[scale=0.85, thick,
				boson/.style={decorate, decoration={snake, segment length=5pt, amplitude=1.5pt}, draw=red!80!black, thick},
				fermion/.style={draw=blue!80!black, thick, postaction={decorate}, decoration={markings, mark=at position 0.55 with {\arrow{Stealth[scale=1.2]}}}},
				vertex/.style={circle, fill=black, inner sep=1.5pt}
				]
				
				\draw[fermion] (-1.8, 1.0) node[left, text=black] {$\bm{k}$} -- (0, 1.0);
				\draw[fermion] (0, 1.0) -- (1.8, 1.0) node[right, text=black] {$\bm{k}'$};
				\draw[fermion] (-1.8, -1.0) node[left, text=black] {$-\bm{k}$} -- (0, -1.0);
				\draw[fermion] (0, -1.0) -- (1.8, -1.0) node[right, text=black] {$-\bm{k}'$};
				
				\draw[boson] (0, 1.0) -- (0, -1.0) node[midway, right=4pt, text=red!80!black] {$\chi(\bm{q}, \omega)$};
				
				\node[vertex, label=above:{\textcolor{blue}{$\Gamma_{\text{eff}} \propto \sin\theta$}}] at (0, 1.0) {};
				\node[vertex, label=below:{\textcolor{blue}{$\Gamma_{\text{eff}} \propto \sin\theta$}}] at (0, -1.0) {};
			\end{tikzpicture} 
			&
			\begin{tikzpicture}[scale=0.95]
				\draw[->, thick] (0,0) -- (3.0,0) node[right] {$\theta$};
				\draw[->, thick] (0,0) -- (0,2.5) node[above] {$\lambda$};
				
				\draw[thick] (-0.08, 0.3) -- (0, 0.3);
				\node[left] at (0, 0.3) {$\lambda_0$};
				
				\draw[dashed, red!80!black, thick] (0,1.0) -- (3.0,1.0) node[pos=0.85, below, yshift=-1pt] {$\lambda_c \approx \mu^*$};
				
				\draw[thick, blue!80!black, domain=0:2.7] plot (\x, {0.3 + 0.25*\x*\x});
				\node[blue!80!black, rotate=45] at (1.0, 1.4) {$\lambda-\lambda_0\propto \sin^2\theta$};
				
				\draw[->, thick, gray] (1.67, 1.0) -- (1.67, 0);
				\node[below, gray] at (1.67, 0) {Threshold};
				
				\fill[blue!80!black, opacity=0.15, domain=1.67:2.7, variable=\x] 
				(1.67, 1.0) -- plot ({\x}, {0.3 + 0.25*\x*\x}) -- (2.7, 1.0) -- cycle;
				\node[blue!80!black] at (2.25, 1.3) {$T_c > 0$};
			\end{tikzpicture}
		\end{tabular}
		\caption{(a) Feynman diagram of the interaction mediated by the overdamped interfacial polar fluctuation. Each exchange contains two dynamic Rashba vertices, so the Rashba-activated contribution scales as $\lambda(\theta)-\lambda_0 \propto |\Gamma_{\text{eff}}|^2 \propto \sin^2\theta$. (b) The dashed line denotes the schematic pairing threshold $\lambda_c \approx \mu^*$, where $\mu^*$ is the effective Coulomb pseudopotential. Above threshold, the nonlinear dependence of $T_c$ on $\lambda$ maps the quadratic angular coupling onto an approximately linear $T_c(\theta)$ over the experimental range.}
		\label{fig:Theory}
	\end{figure}
	
\begin{figure}[tb]
		\centering
		\includegraphics[width=\columnwidth]{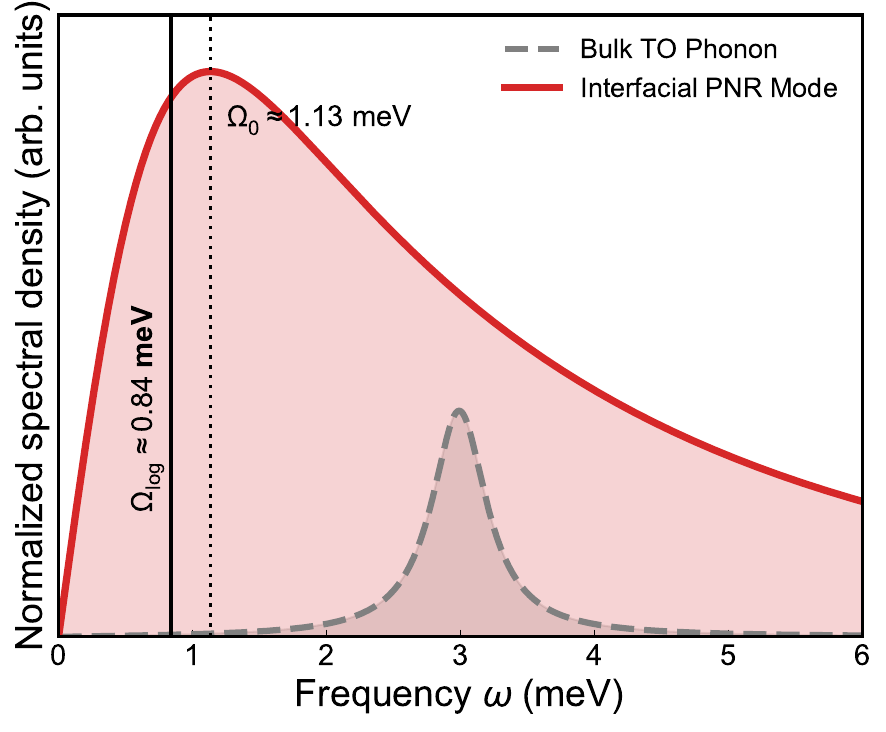}
		\caption{Normalized model spectral densities. The red Brownian line shape $B(\omega)$ represents the effective interfacial spectrum used in the calculation, with orientation entering only through $\alpha^2F(\omega;\theta)=\lambda(\theta)B(\omega)$; the gray dashed curve schematically represents the bulk-KTO TO mode. For the assumed $\Omega_B=3$ meV, the adopted representative value $\Omega_{\log}=0.84$ meV gives a model peak at $\Omega_0\approx1.13$ meV.}
		\label{fig:spectral}
	\end{figure}

	Within Migdal-Eliashberg theory \cite{Eliashberg1960, Carbotte1990}, pairing is encoded in the electron-boson spectral function $\alpha^2F(\omega)$, obtained by averaging the dissipative part of the effective interaction over the Fermi surface:
	\begin{equation} \label{eq:a2F}
		\alpha^2F(\omega) = N(0) \left\langle \frac{1}{\pi} \operatorname{Im} V_{\text{eff}}(\bm{k}, \bm{k}', \omega) \right\rangle_{\text{FS}},
	\end{equation}
	where $N(0)$ is the Fermi-level density of states. We represent this pairing-weighted average by the broad Brownian model spectrum shown in Fig.~\ref{fig:spectral}. Its line shape is assumed, and we adopt the representative logarithmic moment $\Omega_{\log}=0.84$ meV from the correlated near-threshold Allen--Dynes fit family rather than identifying it with the peak position \cite{SupplementalMaterial}. The soft polar modes and low-temperature polar correlations of bulk KTO motivate this choice \cite{Shirane1967, Ichikawa2005, Aktas2014}. Direct spectroscopy of the interfacial polar response would test the required low-energy spectral weight.
	
	In the single-spectrum approximation used here, orientation changes only the overall coupling amplitude:
	\[
		\alpha^2F(\omega;\theta)=\lambda(\theta)B(\omega),
		\qquad
		2\int_0^\infty \frac{B(\omega)}{\omega}\,d\omega=1,
	\]
	The normalization makes $\lambda(\theta)$ the dimensionless zero-frequency coupling, which we parameterize as
	\begin{equation} \label{eq:lambda_total}
		\lambda(\theta) = \lambda_0 + C \sin^2\theta.
	\end{equation}
	Here $\lambda_0$ is the effective orientation-independent baseline and $C\sin^2\theta$ is the Rashba-activated contribution. Both are assigned the same Brownian spectrum; a distinct conventional-phonon contribution would require a separate kernel.
	
	Within the scalar isotropic approximation, $T_c$ is determined from the linearized Matsubara--Eliashberg equations \cite{Eliashberg1960}. For compactness, we write $Z_n(\theta)\equiv Z(i\omega_n;\theta)$, $\Delta_n(\theta)\equiv\Delta(i\omega_n;\theta)$, and $\lambda_{n-m}(\theta)\equiv\lambda(i\omega_n-i\omega_m;\theta)$:
	\begin{align}
		 Z_n(\theta) &=  1 + \frac{\pi k_B T}{\omega_n}
		\sum_m \lambda_{n-m}(\theta)\operatorname{sgn}(\omega_m), \label{eq:Z_ME} \\
		 \Delta_n(\theta) &=  \frac{\pi k_B T}{Z_n(\theta)}
		\sum_m\left[\lambda_{n-m}(\theta)-\mu_m^*\right]
		\frac{\Delta_m(\theta)}{|\omega_m|}. \label{eq:Delta_ME}
	\end{align}
	Here $\omega_n=\pi k_B T(2n+1)$, and $\mu_m^*\equiv\mu^*\Theta(\omega_c-|\omega_m|)$ applies the effective isotropic Coulomb pseudopotential below the cutoff $\omega_c=30$ meV; $\Theta$ denotes the Heaviside step function \cite{Morel1962}. For the bosonic transfer frequency $\nu=\omega_n-\omega_m$, the corresponding effective pairing kernel is
	\begin{align*}
		\lambda(i\nu;\theta)
		&=2\int_0^\infty d\omega\,
		\frac{\omega\,\alpha^2F(\omega;\theta)}{\omega^2+\nu^2}
		=\lambda(\theta)K(i\nu), \\
		K(i\nu)
		&=\frac{\Omega_B^2}{\Omega_B^2+\gamma|\nu|+\nu^2}.
	\end{align*}
	Since $K(0)=1$, $\lambda(0;\theta)=\lambda(\theta)$: Eq.~\eqref{eq:lambda_total} gives the zero-frequency coupling, whereas Eqs.~\eqref{eq:Z_ME} and \eqref{eq:Delta_ME} retain its frequency dependence. We identify $T_c$ as the temperature at which the largest eigenvalue of the linearized gap kernel reaches unity.
	
	To relate the effective spectral scale and coupling to $T_c$, we use the Allen-Dynes interpolation formula \cite{AllenDynes}:
	\begin{equation} \label{eq:AllenDynes}
		k_B T_c = \frac{\Omega_{\log}}{1.2} f_1 f_2 \exp\left[ -\frac{1.04(1+\lambda)}{\lambda - \mu^*(1+0.62\lambda)} \right].
	\end{equation}
	Here $f_1$ and $f_2$ are the phenomenological Allen-Dynes strong-coupling and spectral-shape corrections; they are not derived from the Brownian model and do not enter the numerical Eliashberg calculation. Along the fitted Allen-Dynes curve from (100) to (111), $f_1$ varies from $1.01$ to $1.13$, while $f_2$ varies from $1.00$ to $1.09$. Because Eq.~\eqref{eq:AllenDynes} was calibrated for conventional, well-peaked phonon spectra, its application to the broad overdamped spectrum is an extrapolation. We adopt $\Omega_{\log}=0.84$ meV as a representative point within the correlated near-threshold Allen--Dynes fit family \cite{SupplementalMaterial}. Holding this spectrum and $\mu^*=0.13$ fixed, the coupling parameters are then fitted separately with Allen-Dynes and the numerical Eliashberg equations. At this fixed spectrum, the numerical Eliashberg calculation assesses the angular mapping with $(\lambda_0,C)$ refitted rather than held at their Allen--Dynes values; it does not independently determine $\Omega_{\log}$.
	
\begin{figure}[tb]
		\centering
		\includegraphics[width=\columnwidth]{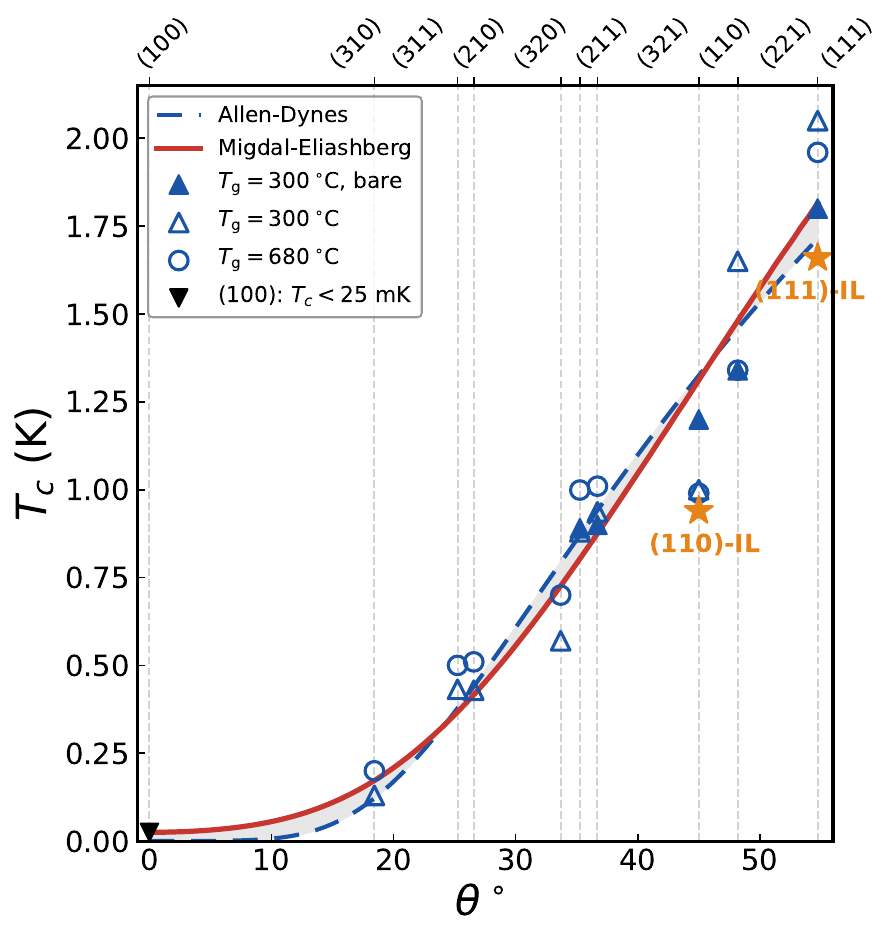}
		\caption{\textbf{Orientation dependence of $T_c$.} Symbols show a-LAO/KTO data \cite{Cao2026} and orange stars show ionic-liquid-gated KTO data \cite{Ren2022}; the (100) marker is a $25$ mK upper bound. Dashed blue and solid red curves are separate Allen--Dynes and numerical Eliashberg fits to the detected a-LAO/KTO transitions using the same spectrum. The shading shows their difference, not a statistical confidence band.}
		\label{fig:Tc_scaling}
	\end{figure}

	With $\Omega_B=3$ meV assumed from the bulk TO scale, the representative $\Omega_{\log}=0.84$ meV determines $\gamma\approx8.8$ meV and gives a model peak at $\Omega_0\approx1.13$ meV. Using this shared spectrum and $\mu^*=0.13$, separate fits to the 23 detected a-LAO/KTO transition temperatures \cite{Cao2026}, subject to the pristine (100) bound $T_c\leq25$ mK, give $(\lambda_0,C)=(0.27,2.87)$ from Allen--Dynes and $(0.39,2.00)$ from numerical Eliashberg [Fig.~\ref{fig:Tc_scaling}]. The different coupling pairs are method-dependent effective parameters. Both fits support the reduced scenario as a viable interpretation of $T_c(\theta)$; however, the comparable residual of a thresholded quadratic shows that fit quality alone does not identify the pairing mechanism \cite{SupplementalMaterial}.
	
	\textbf{Threshold Bounds and Quasi-Linearity.---} The total coupling must exceed a formal pairing threshold $\lambda_c\approx\mu^*$ [Fig.~\ref{fig:Theory}(b)], distinct from the experimental visibility boundary. For the shared spectrum, define $\lambda_{25\,\mathrm{mK}}$ by $T_c(\lambda_{25\,\mathrm{mK}})=25$ mK; it equals $0.418$ in Allen-Dynes and $0.390$ numerically. The absence of a transition at the pristine (100) interface therefore requires $\lambda_0\leq\lambda_{25\,\mathrm{mK}}$. The fitted values $0.27$ and $0.39$, respectively, satisfy this bound, with the numerical fit saturating it. Above $\lambda_c$, the nonlinear $T_c(\lambda)$ relation maps the $\sin^2\theta$ coupling onto an extended quasi-linear $T_c(\theta)$.
	
	\textbf{Tunability of the Threshold.---} The fitted parameters depend on $\mu^*$; the fixed-spectrum estimates below isolate its effect on the (100) threshold \cite{SupplementalMaterial}. Electrostatic boundary conditions can modify the interfacial field, PNR correlation length, and effective Coulomb pseudopotential, thereby shifting $\lambda_0$ relative to $\lambda_c$. To isolate the Coulomb contribution, we hold the shared spectrum and the boundary-saturating numerical baseline $\lambda_0=0.390$ fixed. Reducing $\mu^*$ from $0.13$ to $0.10$ then raises the calculated $T_c$ at (100) from $25$ to $\sim40$ mK, and to $\sim55$ mK at $\mu^*=0.08$ and $\sim104$ mK at $\mu^*=0.04$. These are conditional sensitivity estimates for a-LAO/KTO, because screening may also modify the polar spectrum and electron-boson coupling. A controlled screening-induced onset of superconductivity at pristine a-LAO/KTO(100) would test this threshold scenario. Nonvolatile switching of the out-of-plane polarization changes the local electrostatic environment and is accompanied by bistability and $T_c$ enhancement \cite{Zhang2026, Dong2026}. Related electrostatic differences may also contribute to the modest intercept offset between a-LAO-capped and ionic-liquid-gated platforms, which exhibit similar slopes \cite{Cao2026, Ren2022}.
	
	\textbf{Discussion and Conclusion.---} The reduced scalar isotropic model isolates how a geometrically controlled dynamic Rashba vertex strengthens pairing mediated by soft interfacial polar modes. A complete microscopic treatment would additionally require the full multiband $t_{2g}$ structure, facet-dependent confinement, and Fermi-surface anisotropy.
	
	The fitted couplings place the (111) interface in the strong-coupling regime: $\lambda\approx2.18$ for the Allen-Dynes fit and $\lambda\approx1.72$ for the numerical Eliashberg fit; the latter corresponds to a quasiparticle residue $1/(1+\lambda)\approx0.37$. Moreover, because an overdamped mode has no unique characteristic frequency $\Omega$, the conventional Migdal parameter $\lambda\Omega/E_F$ does not by itself control vertex corrections. Assessing these effects requires a treatment beyond the present isotropic equations \cite{SupplementalMaterial}.

	The larger atomic spin-orbit coupling of Ta $5d$ orbitals ($\xi_{5d}\sim400$ meV versus $\xi_{3d}\sim20$ meV for Ti $3d$ orbitals \cite{King2012, Zhong2013}) offers a parametric rationale for the stronger orientation dependence and higher $T_c$ of KTO interfaces relative to STO. In the reduced model, stronger SOC can enhance the vertex amplitude $g_0$ in Eq.~\eqref{eq:g_theta}. Because boson exchange involves two such vertices, $C\propto g_0^2$, while its magnitude also depends on $N(0)$, $k_F$, the polar spectrum, and the helicity projection; $\lambda_0$ remains a separate orientation-independent baseline. A quantitative KTO--STO comparison would require material-specific orbital composition, confinement, hybridization, carrier density, and polar spectra for both interfaces.
	
	Several experiments offer direct tests of the proposed polar-Rashba scenario and routes to distinguish it from purely electronic pairing theories \cite{Palle2024} and isotropic bulk paraelectric-phonon models \cite{Gastiasoro2020}. Interface-sensitive spectroscopy can search for low-energy polar spectral weight at the interface. Orientation-resolved tunneling can probe the facet dependence of the mixed-parity gap structure enabled by dynamic Rashba coupling \cite{Kozii2015, SupplementalMaterial}. Oxygen-isotope substitution can probe the role of Ta--O dynamics through the isotope coefficient $\alpha=-d\ln T_c/d\ln M_{\mathrm O}$, with $\alpha=0.5$ as the harmonic benchmark; deviations would indicate physics beyond that benchmark but would not uniquely identify its origin. Finally, electrostatic tuning can test the conditional prediction that enhanced screening drives the pristine a-LAO/KTO(100) interface above the present $25$-mK visibility limit.

	In conclusion, we have developed a minimal theoretical framework for a-LAO/KTO interfacial superconductivity in which a geometrically controlled dynamic Rashba vertex couples electrons to overdamped polar modes, linking interface orientation to the observed quasi-linear $T_c(\theta)$. The agreement supports this polar pairing scenario as a viable candidate mechanism for orientation-dependent superconductivity at KTO interfaces.
	
	\let\oldaddcontentsline\addcontentsline
	\renewcommand{\addcontentsline}[3]{}
	
	\begin{acknowledgments}
		This work was supported by the National Key R\&D Program of China (Grant No. 2022YFA1403403) and the National Natural Science Foundation of China (Grant Nos. 12274441 and 12534004). The author thanks M.~Zhang and Y.~Xie for stimulating discussions and long-term collaboration on KTaO$_3$ interfaces.
	\end{acknowledgments}
	
	\section*{Data and Code Availability}
	The experimental data and Python scripts used for the Allen--Dynes and numerical Eliashberg calculations, fitting, and generation of Figs.~3, 4, and S2 are publicly available in the GitHub repository \texttt{KTO-Geometry-SOC}~\cite{KTO_Geometry_SOC}. The repository also archives the AI-assistance transcripts and figure-design prompts retained during manuscript preparation.

	\bibliography{references}
	
	\let\addcontentsline\oldaddcontentsline
	
	\clearpage
	\onecolumngrid
	\begin{center}
		\textbf{\large Supplementary Material for ``Geometric Rashba Control of Polar Pairing at a-LaAlO$_3$/KTaO$_3$ Interfaces''}
	\end{center}
	\vspace{1em}
	
	\tableofcontents
	\vspace{2em}
	\clearpage
	
	\setcounter{equation}{0}
	\setcounter{figure}{0}
	\setcounter{table}{0}
	\setcounter{page}{1}
	\renewcommand{\thepage}{S\arabic{page}}
	\renewcommand{\theequation}{S\arabic{equation}}
	\renewcommand{\thefigure}{S\arabic{figure}}
	\renewcommand{\thetable}{S\arabic{table}}
	\renewcommand{\theHequation}{S\arabic{equation}}
	\renewcommand{\theHfigure}{S\arabic{figure}}
	\renewcommand{\theHtable}{S\arabic{table}}
	
	\section{Supplementary Note 1: Reduced Orbital Rotation and the Angular Ansatz}
	
	For a cubic crystal, symmetry-related facets can be represented in the fundamental domain $h\ge k\ge l\ge0$. We use (100), with normal $\hat{x}$, as the reference orientation and define
	\[
	\theta=\arccos\!\left(\frac{h}{\sqrt{h^2+k^2+l^2}}\right),
	\qquad 0\leq\theta\leq\arccos(1/\sqrt{3}).
	\]
	This angle does not uniquely specify a facet. For example, (110) and (543) both have $\theta=45^\circ$ but are not symmetry-equivalent. A general normal also requires an azimuth $\phi$, with $\hat n=(\cos\theta,\sin\theta\cos\phi,\sin\theta\sin\phi)$ in the cubic frame. The reduction to $g(\theta)$ neglects this additional dependence, as well as possible facet dependence of confinement and orbital composition.
	
	To motivate one geometrically activated channel, restrict the general normal to the illustrative $x$--$z$ plane ($\phi=\pi/2$) and consider a representative Ta $|d_{yz}\rangle$ orbital. This is not a construction of the electronic eigenstates at arbitrary $(hkl)$ interfaces. In local coordinates with $x'$ along the tilted interface normal, take
	\begin{align}
		 x &= x'\cos\theta-z'\sin\theta, \\
		 y &= y', \\
		 z &= x'\sin\theta+z'\cos\theta.
	\end{align}
	Then $yz=\cos\theta\,y'z'+\sin\theta\,x'y'$, so that, within this selected rotation,
	\begin{equation}
		|d_{yz}\rangle=\cos\theta|d_{y'z'}\rangle+\sin\theta|d_{x'y'}\rangle.\label{eq:dyz}
	\end{equation}
	The representative $d_{yz}$-derived subband is motivated by its heavy normal mass and low confinement energy near (100); its continued dominance under tilt is not derived. Nor is the cubic $t_{2g}$ subspace closed under a general spatial rotation: the simple two-orbital identity above applies to the chosen rotation, not to arbitrary facets.
	
	In a centered atomic-orbital dipole approximation, a normal electric-field fluctuation gives
	\begin{align}
	\langle p_{y'}|e\delta E_{x'}x'|d_{yz}\rangle
	&=e\delta E_{x'}\big[\cos\theta\langle p_{y'}|x'|d_{y'z'}\rangle
	+\sin\theta\langle p_{y'}|x'|d_{x'y'}\rangle\big] \\
	&=e\delta E_{x'}\sin\theta\langle p_{y'}|x'|d_{x'y'}\rangle.
	\end{align}
	The first integral vanishes by $x'$ parity, while the second is allowed. This supplies a $\sin\theta$ factor for the selected dipole channel. Mapping that factor onto an effective low-energy Rashba vertex further assumes that the relevant hybridization, SOC matrix elements, and intermediate-state energy denominators can be absorbed into an orientation-independent coefficient $g_0$.
	
	We therefore adopt $g(\theta)=g_0\sin\theta$ as a reduced angular ansatz motivated by this channel. It is not a point-group prohibition of all dynamic Rashba coupling at (100), where inversion symmetry is already broken by the interface. Additional orbitals, azimuthal dependence, and confinement-induced mixing can produce other contributions. The baseline $\lambda_0$ parameterizes unresolved orientation-independent attraction; whether other channels can be represented by this baseline requires a more complete microscopic treatment.
	
	\section{Supplementary Note 2: Symmetry Form of the Dynamic Rashba Vertex and Helicity Projection}
	The relativistic SOC operator motivates the symmetry form of a coupling between normal polar fluctuations and interfacial electrons. It does not by itself calculate the effective coefficient in a multiorbital solid. Here we construct that allowed interaction and describe the assumptions used to reduce it to a scalar kernel.
	
	\textbf{1. Fundamental SOC and the Out-of-Plane Polar Fluctuation}
	
	The fundamental relativistic spin-orbit interaction for an electron moving in an electric field $\mathbf{E}$ is given by:
	\begin{equation}
		\mathcal{H}_{\text{SO}} \propto (\mathbf{E} \times \mathbf{k}) \cdot \boldsymbol{\sigma}.
	\end{equation}
	Within the reduced model, we retain only the normal components of the static confinement field, $\mathbf{E}_{\text{static}} = E_0 \hat{n}$, and the dynamic PNR amplitude fluctuation, $\delta\mathbf{E} = \delta E_z \hat{n}$. Here $z$ denotes the local normal direction and $\delta E_z$ is proportional to the polar displacement field $P_z(\mathbf{q})$. The modeled field is therefore $\mathbf{E} = (E_0 + \delta E_z) \hat{n}$; possible in-plane components lie outside this scalar construction.
	
	Substituting this total field into the fundamental SOC Hamiltonian yields:
	\begin{equation}
		\mathcal{H}_{\text{SO}} \propto (E_0 + \delta E_z) (\hat{n} \times \mathbf{k}) \cdot \boldsymbol{\sigma} = (E_0 + \delta E_z) (\mathbf{k} \times \boldsymbol{\sigma}) \cdot \hat{n}.
	\end{equation}
	Choosing local in-plane axes $(x,y)$ perpendicular to $\hat{n}$, the cross product expands as $(\mathbf{k} \times \boldsymbol{\sigma}) \cdot \hat{n} = (k_x \sigma_y - k_y \sigma_x)$. The Hamiltonian naturally separates into a static Rashba term and a dynamic electron-boson interaction:
	\begin{equation}
		\mathcal{H}_{\text{SO}} = \alpha_R (\mathbf{k} \times \boldsymbol{\sigma}) \cdot \hat{n}
		+ g(\theta) \sum_{\mathbf{k},\mathbf{q}} P_z(\mathbf{q})
		\psi^\dagger_{\mathbf{k}+\mathbf{q}/2}
		\left[ (\mathbf{k} \times \boldsymbol{\sigma}) \cdot \hat{n} \right]
		\psi_{\mathbf{k}-\mathbf{q}/2}.
	\end{equation}
	At the effective-band level, the dynamic coefficient involves the response $\partial\alpha_R/\partial P_z$ about the equilibrium interfacial state. It need not share the angular dependence of the full static $\alpha_R$, which can contain contributions from other orbital channels. We use the selected dipole channel of Supplementary Note~1 to motivate $g(\theta)=g_0\sin\theta$, without imposing $\alpha_R(\theta)\propto\sin\theta$ on the static band structure.
	
	\textbf{2. Projection onto the Helicity Basis}
	
	The dynamic interaction $H_{\text{int}} \propto (\mathbf{k} \times \boldsymbol{\sigma}) \cdot \hat{n}$ is matrix-valued and momentum-dependent. To construct a scalar-dominant Eliashberg pairing strength $\lambda$, we project this interaction onto the eigenstates of the unperturbed Fermi surface.
	
	The static non-interacting Hamiltonian is given by $H_0 = \frac{\hbar^2 k^2}{2m^*} + \alpha_R \hat{\Lambda}_{\mathbf{k}}$, where we define the helicity operator:
	\begin{equation}
		\hat{\Lambda}_{\mathbf{k}} = (\mathbf{k} \times \boldsymbol{\sigma}) \cdot \hat{n} = k_x \sigma_y - k_y \sigma_x.
	\end{equation}
	The eigenstates of $H_0$ are the Rashba-split helicity bands $|\pm, \mathbf{k}\rangle$. Because the Rashba term in $H_0$ is proportional to $\hat{\Lambda}_{\mathbf{k}}$, the helicity bands are eigenstates of $\hat{\Lambda}_{\mathbf{k}}$ with eigenvalues proportional to the magnitude of the momentum:
	\begin{equation}
		\hat{\Lambda}_{\mathbf{k}} |\pm, \mathbf{k}\rangle = \pm |\mathbf{k}| |\pm, \mathbf{k}\rangle.
	\end{equation}
	This dictates that the electron spins lie entirely in the $xy$-plane, tangentially locked perpendicular to $\mathbf{k}$.
	
	\begin{figure}[tb]
		\centering
		\begin{tikzpicture}[scale=1.2,
			spinarrow/.style={-Stealth, thick, red!80!black},
			spinarrow2/.style={-Stealth, thick, blue!80!black},
			boson/.style={decorate, decoration={snake, segment length=4pt, amplitude=1.2pt}, draw=orange!90!black, thick, -Stealth}
			]
			
			\draw[->, gray!70] (-3, 0) -- (3, 0) node[right, text=black] {$k_x$};
			\draw[->, gray!70] (0, -3) -- (0, 3) node[above, text=black] {$k_y$};
			
			\draw[thick, blue!80!black, dashed] (0,0) circle (1.2);
			\node[text=blue!80!black] at (-0.5, 0.5) {$|+, \mathbf{k}\rangle$};
			
			\draw[thick, red!80!black] (0,0) circle (2.2);
			\node[text=red!80!black] at (-1.5, 1.5) {$|-, \mathbf{k}\rangle$};
			
			\draw[spinarrow2] (1.2, 0) -- (1.2, 0.4);
			\draw[spinarrow2] (0, 1.2) -- (-0.4, 1.2);
			\draw[spinarrow2] (-1.2, 0) -- (-1.2, -0.4);
			\draw[spinarrow2] (0, -1.2) -- (0.4, -1.2);
			
			\draw[spinarrow] (2.2, 0) -- (2.2, -0.4);
			\draw[spinarrow] (0, 2.2) -- (0.4, 2.2);
			\draw[spinarrow] (-2.2, 0) -- (-2.2, 0.4);
			\draw[spinarrow] (0, -2.2) -- (-0.4, -2.2);
			
			\fill[black] (1.9, 1.1) circle (2pt) node[right] {$\mathbf{k}$};
			\fill[black] (1.1, 1.9) circle (2pt) node[above right] {$\mathbf{k}'$};
			
			\draw[boson] (1.9, 1.1) to[bend left=20] node[midway, below left, text=orange!90!black] {$P_z(\mathbf{q})$} (1.1, 1.9);
			
			\draw[-Stealth, very thick, red!80!black] (1.9, 1.1) -- (2.3, 0.4) node[right] {$\mathbf{S}_{\mathbf{k}}$};
			\draw[-Stealth, very thick, red!80!black] (1.1, 1.9) -- (1.8, 1.5) node[right] {$\mathbf{S}_{\mathbf{k}'}$};
			
		\end{tikzpicture}
			\caption{Illustration of the helicity projection on the Rashba-split Fermi surfaces. The static Rashba SOC splits the bands into an inner $|+\rangle$ and outer $|-\rangle$ helicity contour, where spins (colored arrows) are locked tangentially to the momentum. Because the dynamic electron-boson interaction operator $\hat{\Lambda}_{\mathbf{k}}$ has the same matrix form as the static Rashba operator, projecting the interaction onto the dominant, low-energy outer band $|-, \mathbf{k}\rangle$ diagonalizes the vertex. For soft relaxational bosons mediating forward-scattering ($\mathbf{k} \to \mathbf{k}' \approx \mathbf{k}$), the spinor overlap approaches unity, reducing the matrix-valued dynamic Rashba vertex to an effective scalar intra-band kernel.}
		\label{fig:Helicity}
	\end{figure}
	
	We now project the dynamic interaction vertex $\hat{\Gamma}(\bar{\mathbf{k}}) = g(\theta) \hat{\Lambda}_{\bar{\mathbf{k}}}$ onto the lowest helicity band $|-, \mathbf{k}\rangle$, which is assumed to host the dominant low-energy 2DEG states at the LAO/KTO interface. Let $\bar{\mathbf{k}}=(\mathbf{k}+\mathbf{k}')/2$, with $|\mathbf{k}|=|\mathbf{k}'|=k_F$. For either helicity $s=\pm$, the intra-band matrix element on a circular Rashba contour is
	\begin{equation}
		\langle s,\mathbf{k}'|\hat{\Gamma}(\bar{\mathbf{k}})|s,\mathbf{k}\rangle
		=s\,g(\theta)k_F\langle s,\mathbf{k}'|s,\mathbf{k}\rangle,
	\end{equation}
	where the magnitude of the spinor overlap is $|\cos[(\phi_{\mathbf{k}'}-\phi_{\mathbf{k}})/2]|$. For the assumed small-$q$ forward-scattering sector, this overlap approaches unity. Selecting the lower helicity sheet and replacing the residual momentum dependence by a representative Fermi-surface value then gives the effective scalar amplitude $\Gamma_{\text{eff}}\approx-g(\theta)k_F$.
	
	\textbf{3. Reduction to a Scalar Eliashberg Kernel}
	
	Within the Eliashberg formalism, the pairing strength $\lambda$ is obtained by integrating out the bosonic field. The effective four-fermion interaction $V_{\text{eff}}$ mediated by the boson involves the product of two vertices and the boson propagator $\chi(\mathbf{q}, \omega)$:
	\begin{equation}
		V_{\text{eff}} \propto |\Gamma_{\text{eff}}|^2 \chi(\mathbf{q}, \omega).
	\end{equation}
	Within the selected helicity sheet, the sign of the forward-limit projected vertex drops out of its squared magnitude:
	\begin{equation}
		|\langle \Gamma_{\text{eff}} \rangle|^2 \approx k_F^2 g^2(\theta) \propto \sin^2\theta.
	\end{equation}
	This projection simplifies two critical points:
	\begin{enumerate}
		\item In the limit of dominant intra-band, small-$q$ forward scattering, it motivates the reduction of the full mixed-parity Nambu-Gorkov equations to the standard, isotropic Eliashberg equations utilized in Supplementary Note 4.
		\item Because the effective pairing strength $\lambda$ is proportional to the square of this scalar vertex ($\lambda \propto |\Gamma_{\text{eff}}|^2$), the Eliashberg coupling inherits the geometric scaling: $\lambda(\theta) = \lambda_0 + C \sin^2\theta$. This provides the analytical input required for the macroscopic Allen-Dynes $T_c$ evaluation.
	\end{enumerate}
	
	\textbf{4. Caveat Regarding the Spinor Overlap Factor}
	
		We note that the full momentum-dependent projection of the Rashba vertex between different momentum states $\mathbf{k}$ and $\mathbf{k}'$ introduces a complex angular spinor overlap whose magnitude is proportional to $\cos[(\phi_{\mathbf{k}} - \phi_{\mathbf{k}'})/2]$ in the forward sector. This angular dependence contributes to the momentum-dependent pairing structure needed to reconstruct the singlet--triplet content. The relative admixture also depends on the interaction and the gaps on both helicity sheets, as discussed in Supplementary Note~7.
	
		However, explicitly retaining this momentum-selective overlap converts the Eliashberg gap equations into a computationally intensive multi-channel anisotropic problem. By replacing this factor with an isotropic Fermi-surface average, our scalar reduction operates as an effective model. This reduction parameterizes the integrated pairing scale while sacrificing the momentum-dependent gap structure. Its quantitative accuracy, including the effects of the other helicity band, is not established by the scalar calculation. A momentum-dependent multiband treatment is required to test the projection and determine singlet--triplet mixing.
	
	\section{Supplementary Note 3: Effective Brownian Response and the Isotropic Spectral Approximation}
	We model a soft finite-momentum sector of the out-of-plane PNR response with a second-order damped oscillator. This phenomenological equation incorporates inertia and damping; it is not a microscopic derivation of the interfacial spectrum. In particular, it does not include the nonlocal electrostatic contribution responsible for stiffening the uniform mode. The inertial term makes the second Eliashberg moment finite, unlike a purely first-order relaxator.
	
	Define the spatial dispersion factor as $A_q \equiv 1 + \xi_P^2 q^2$. For a polar lattice mode with inertia, the dynamics of the polar displacement field $P_{\bm{q}}(t)$ driven by a conjugate source field $h_{\bm{q}}(t)$ are governed by a damped wave equation:
	\begin{equation}
		\partial_t^2 P_{\bm{q}}(t) + \gamma \partial_t P_{\bm{q}}(t) + \Omega_B^2 A_q P_{\bm{q}}(t) = \Omega_B^2 \chi_0 h_{\bm{q}}(t),
	\end{equation}
	where $\Omega_B$ is the oscillator frequency parameter, $\gamma$ the phenomenological damping rate, and $\chi_0$ a static susceptibility scale for this soft-sector model.
	
	Using the standard Fourier convention:
	\begin{equation}
		P_{\bm{q}}(t) = \int \frac{d\omega}{2\pi} e^{-i\omega t} P_{\bm{q}}(\omega),
	\end{equation}
	the time derivatives map to the frequency domain as $\partial_t \to -i\omega$ and $\partial_t^2 \to -\omega^2$. Substituting these into the equation of motion gives:
	\begin{equation}
		(-\omega^2 - i\gamma \omega + \Omega_B^2 A_q) P_{\bm{q}}(\omega) = \Omega_B^2 \chi_0 h_{\bm{q}}(\omega).
	\end{equation}
	
	The retarded dynamical susceptibility is defined as the linear response $\chi(\bm{q}, \omega) \equiv P_{\bm{q}}(\omega)/h_{\bm{q}}(\omega)$. Dividing both sides by the source field yields the form used in Eq.~(1) of the main text:
	\begin{equation}
		\chi(\bm{q}, \omega) = \frac{\chi_0 \Omega_B^2}{\Omega_B^2 A_q - \omega^2 - i\gamma \omega} = \frac{\chi_0 \Omega_B^2}{\Omega_B^2(1 + \xi_P^2 q^2) - \omega^2 - i \gamma \omega}.
	\end{equation}
	The oscillator at a given momentum is overdamped when
	$\gamma>2\Omega_B\sqrt{A_q}$.
	
	In the low-frequency regime of an overdamped oscillator, $|\omega|\ll\gamma$, the inertial $-\omega^2$ term is subleading. Dividing numerator and denominator by $\Omega_B^2$ then yields
	\begin{equation}
		\chi(\bm{q}, \omega) \approx \frac{\chi_0}{A_q - i\omega (\gamma/\Omega_B^2)}
		=\frac{\chi_0/A_q}{1-i\omega/\Omega_R(q)},
	\end{equation}
	where $\Omega_R(q)\equiv\Omega_B^2A_q/\gamma$ is the momentum-dependent relaxation rate. Thus, at low energies, the mode behaves as an overdamped relaxator and provides broad low-energy spectral weight.
	
	However, at high energies (the UV limit), the full Brownian equation retains the inertial $-\omega^2$ term in the denominator. Thus $\chi(\omega)\propto\omega^{-2}$ while its dissipative part decays as $\operatorname{Im}\chi(\omega)\propto\omega^{-3}$. Because $\alpha^2F(\omega)$ is proportional to $\operatorname{Im}\chi(\omega)$, this fast decay regularizes the Eliashberg integral and keeps the second moment $\bar{\omega}_2$ finite.
	
	\textbf{Separate reduction to a single effective spectrum.} For the $q$-dependent response above, even a scalar, frequency-independent vertex generally gives a normalized kernel of the form
	\[
	K_{\mathrm{av}}(i\nu)=
	\frac{\displaystyle\left\langle |\Gamma_{\bm k\bm k'}|^2
	\frac{\Omega_B^2}{\Omega_B^2A_q+\gamma|\nu|+\nu^2}\right\rangle_{\mathrm{FS}}}
	{\displaystyle\left\langle |\Gamma_{\bm k\bm k'}|^2/A_q\right\rangle_{\mathrm{FS}}},
	\qquad \bm q=\bm k-\bm k'.
	\]
	A superposition over $A_q$ need not retain a single Brownian line shape. The numerical calculations do not evaluate this average or determine the relevant momentum window. Instead, they approximate the integrated response by
	\[
	B(\omega)=\frac{1}{\pi}\frac{\Omega_B^2\gamma\omega}{(\Omega_B^2-\omega^2)^2+\gamma^2\omega^2},
	\qquad
	K(i\nu)=\frac{\Omega_B^2}{\Omega_B^2+\gamma|\nu|+\nu^2}.
	\]
	The line shape satisfies $2\int_0^\infty B(\omega)\,d\omega/\omega=1$ and generates the stated kernel through $K(i\nu)=2\int_0^\infty \omega B(\omega)\,d\omega/(\omega^2+\nu^2)$.
	Here the same symbols $\Omega_B$ and $\gamma$ label effective spectral parameters. The form coincides algebraically with setting $A_q=1$, but is not an identification of the pairing boson with the physical uniform mode. Assigning the bulk-inspired value $\Omega_B=3$ meV to this effective spectrum is an additional modeling choice. Establishing its relation to the finite-$q$ polar response requires the dispersion, electrostatic screening, and momentum-dependent vertex, none of which is determined by the present fit.
	
	\section{Supplementary Note 4: Review of Isotropic Eliashberg Equations}
	
	Within the reduced model, the pairing interaction is generated by the exchange of a soft bosonic mode between two electrons. As illustrated by the Feynman diagram in Fig.~2(a) of the main text, a Cooper pair scatters from momentum $(\bm{k}, -\bm{k})$ to $(\bm{k}', -\bm{k}')$ by exchanging the overdamped PNR amplitude fluctuation $\chi(\bm{q}, \omega)$. 
	
	With the pairing interaction approximated as a scalar-dominant kernel via the helicity projection (as detailed in Supplementary Note 2), we separate its angle-dependent amplitude from its frequency dependence as
	\[
		\alpha^2F(\omega;\theta)=\lambda(\theta)B(\omega),
		\qquad
		2\int_0^\infty\frac{B(\omega)}{\omega}\,d\omega=1.
	\]
		The corresponding isotropic Eliashberg equations for the mass renormalization and gap function take the standard form on the Matsubara axis, where $\omega_n=\pi k_BT(2n+1)$, $Z_n(\theta)\equiv Z(i\omega_n;\theta)$, $\Delta_n(\theta)\equiv\Delta(i\omega_n;\theta)$, and $\lambda_{n-m}(\theta)\equiv\lambda(i\omega_n-i\omega_m;\theta)$:
		\begin{align}
			 Z_n(\theta) &=  1 + \frac{\pi k_BT}{\omega_n}
			\sum_m \lambda_{n-m}(\theta)\operatorname{sgn}(\omega_m), \label{eq:Z_app} \\
			 \Delta_n(\theta) &=  \frac{\pi k_BT}{Z_n(\theta)}
			\sum_m\left[\lambda_{n-m}(\theta)-\mu_m^*\right]
			\frac{\Delta_m(\theta)}{|\omega_m|}. \label{eq:Delta_app}
	\end{align}
	Here $\mu_m^*\equiv\mu^*\Theta(\omega_c-|\omega_m|)$ is the Coulomb pseudopotential with cutoff $\omega_c=30$ meV, and $\Theta$ is the Heaviside function. The effective pairing kernel is
	\[
		\lambda(i\nu;\theta)
		=2\int_0^\infty d\omega\,
		\frac{\omega\,\alpha^2F(\omega;\theta)}{\omega^2+\nu^2}
		=\lambda(\theta)K(i\nu),
		\qquad
		K(i\nu)=\frac{\Omega_B^2}{\Omega_B^2+\gamma|\nu|+\nu^2}.
	\]
	Because $K(0)=1$, the zero-frequency kernel equals the pairing amplitude of Eq.~\eqref{eq:lambda_total}: $\lambda(0;\theta)=\lambda(\theta)$. The calculation applies this same normalized Brownian kernel to both the effective angle-independent baseline $\lambda_0$ and the Rashba-activated term $C\sin^2\theta$. A separate conventional-phonon contribution would require an additional spectral kernel. We identify $T_c$ as the temperature at which the largest eigenvalue of the linearized gap kernel reaches unity. The Allen-Dynes calibration is discussed in Supplementary Notes~5 and 6, while the separate numerical Eliashberg refit is described in Supplementary Note~9.
	
	\section{Supplementary Note 5: Allen-Dynes Shape Factors and UV Regularization}
	The Allen-Dynes interpolation augments the McMillan expression with two phenomenological shape factors, $f_1$ and $f_2$:
	\begin{equation}
		k_B T_c = \frac{\Omega_{\log}}{1.2} f_1 f_2 \exp\left[ -\frac{1.04(1+\lambda)}{\lambda - \mu^*(1+0.62\lambda)} \right].
	\end{equation}
	The spectral moments and shape factors entering this expression are
	\begin{align}
		\Omega_{\log} &= \exp\left( \frac{2}{\lambda} \int_0^\infty \frac{d\omega}{\omega} \alpha^2F(\omega) \ln\omega \right), \\
		\bar{\omega}_2 &= \left(\frac{2}{\lambda}\int_0^\infty \omega\alpha^2F(\omega)\,d\omega\right)^{1/2}, \\
		f_1 &= \left[ 1 + \left(\frac{\lambda}{\Lambda_1}\right)^{3/2} \right]^{1/3}, \quad \Lambda_1 = 2.46(1+3.8\mu^*), \\
		f_2 &= 1 + \frac{(\bar{\omega}_2/\Omega_{\log} - 1)\lambda^2}{\lambda^2 + \Lambda_2^2}, \quad \Lambda_2 = 1.82(1+6.3\mu^*) \frac{\bar{\omega}_2}{\Omega_{\log}}.
	\end{align}
	We emphasize that $f_1$ and $f_2$, including the numerical coefficients entering $\Lambda_1$ and $\Lambda_2$, are empirical interpolation factors obtained by Allen and Dynes from fits to numerical Eliashberg solutions for conventional phonon spectra \cite{AllenDynes}; they are not derived from the overdamped Brownian-oscillator model used here. The spectral moments $\Omega_{\log}$ and $\bar{\omega}_2$ are computed from our spectrum.

	For a pure Debye relaxational susceptibility, the high-frequency tail of the spectral function scales as $\alpha^2 F(\omega) \sim 1/\omega$. The integral defining $\bar{\omega}_2^2$ then diverges linearly in the ultraviolet (UV).
	
	However, by adopting the overdamped Brownian oscillator model for the soft PNR mode as derived in Supplementary Note 3, $\chi(\omega)$ drops as $1/\omega^2$ at high frequencies. This yields a spectral function that decays as $\alpha^2 F(\omega) \sim \omega^{-3}$. This rapid decay regularizes the phenomenological spectrum at high frequency, guaranteeing that $\bar{\omega}_2$ is finite; it does not supply a material-specific microscopic cutoff.
	
	\section{Supplementary Note 6: Details of the Allen-Dynes Curve Fitting}
	To connect the reduced framework to the experimental data, we examined the Allen-Dynes fit family with $\Omega_B=3$ meV and $\mu^*=0.13$ fixed, varying the damping and coupling parameters while recomputing the Brownian moments $\Omega_{\log}$ and $\bar{\omega}_2$ at each step. Changes in $\Omega_{\log}$, $\lambda_0$, and $C$ compensate one another, yielding nearly indistinguishable $T_c(\theta)$ curves rather than a unique spectral-scale extraction. For example, imposing the explicit low-baseline prior $0.10\leq\lambda_0\leq0.25$ gives a constrained optimum at $\lambda_0=0.250$, with $\Omega_{\log}=0.882$ meV, $C=2.779$, and $\mathrm{SSR}=0.58936$ K$^2$. At the adopted representative value $\Omega_{\log}=0.840$ meV, refitting the couplings gives $\lambda_0=0.2725$, $C=2.8665$, and $\mathrm{SSR}=0.58890$ K$^2$. The negligible residual difference and the boundary-active first result show that the fit admits a broad near-threshold family, not a unique $\Omega_{\log}$. We therefore use $0.84$ meV as a representative scale. The final curves in Fig.~4 and the contours below hold this shared spectrum fixed and refit only $(\lambda_0,C)$ separately with Allen-Dynes and the numerical Eliashberg equations.
	
	The residual is evaluated on the 23 detected amorphous-LAO/KTO transitions shown by the blue symbols in Fig.~4. The ionic-liquid-gated points (orange stars) are displayed for comparison only, while the pristine (100) measurement is imposed as the upper-limit constraint $T_c\leq25$ mK rather than entered as an artificial $T_c=0$ data point.
	
	The resulting fitting landscape contains strongly correlated parameter directions. In particular, both a near-threshold family of solutions with relatively small $\lambda_0$ and large $C$, and a higher-baseline family with somewhat smaller geometric enhancement, can provide reasonable descriptions of the data. In the main text, we focus on the physically motivated near-threshold family because it is most naturally connected to the physical collapse of superconductivity at the pristine (100) interface and to the interpretation of the geometric Rashba channel as the dominant orientation-dependent enhancement.

	The contours are conditional on $\Omega_B$, $\Omega_{\log}$, and $\mu^*$ being held fixed. They omit uncertainty in the spectral scale inferred from the same transition data and do not represent full joint uncertainty in the model. In addition, intersecting conventional $F$-based regions with an active parameter bound is only an approximate construction, not a calibrated boundary-aware coverage calculation.

\begin{figure}[tb]
		\centering
		\includegraphics[width=\columnwidth]{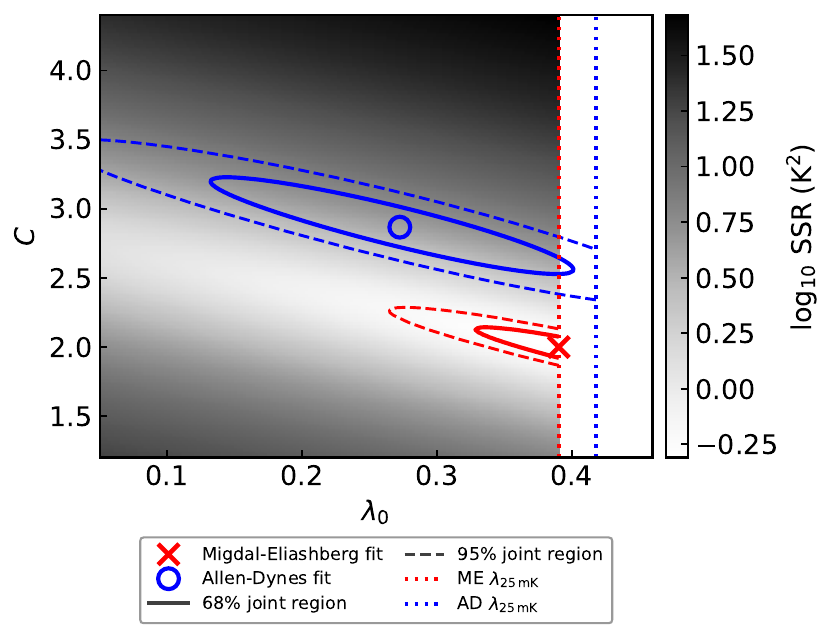}
		\caption{Constrained residual landscape of the $(\lambda_0,C)$ fits at the fixed spectrum of the main text ($\Omega_B=3$ meV, $\Omega_{\log}=0.84$ meV, $\mu^*=0.13$). The residual is evaluated on the 23 detected transitions; the pristine (100) result instead imposes the method-specific bounds $\lambda_0\leq\lambda_{25\,\mathrm{mK}}$. Gray shading shows the numerical Eliashberg residual within its allowed domain. Red and blue contours are the approximate $68\%$ and $95\%$ joint confidence regions for the numerical Eliashberg and Allen-Dynes fits, truncated at the corresponding dotted visibility bounds. The cross and circle mark their constrained minima.}
		\label{fig:fit_covariance}
	\end{figure}

	Figure~\ref{fig:fit_covariance} makes the covariance explicit using the same convention as Fig.~4: the residual contains the 23 detected transitions, while the pristine (100) result restricts each method to $\lambda_0\leq\lambda_{25\,\mathrm{mK}}$ rather than entering as an artificial zero. Because the reported $T_c$ values carry no quoted uncertainties, the contours are approximate joint confidence regions from the $F$-distribution for nonlinear least squares with unknown variance, based on the sum of squared residuals (SSR) over the $N=23$ detected transitions: $\mathrm{SSR}\leq\mathrm{SSR}_{\min}\big[1+\tfrac{p}{N-p}F_{p,N-p}(\alpha)\big]$, with $p=2$, intersected with the method-specific visibility bounds. Each elongated valley shows that a larger baseline trades against a weaker geometric amplitude. Their constrained minima are $(\lambda_0,C)=(0.2725,2.8665)$ with $\mathrm{SSR}=0.58890$ K$^2$ for Allen-Dynes and $(0.3904,1.9999)$ with $\mathrm{SSR}=0.48930$ K$^2$ numerically. For $N=23$ and $p=2$, the corresponding residual standard errors, $s=\sqrt{\mathrm{SSR}/(N-p)}$, are $0.167$ K and $0.153$ K, respectively. The numerical minimum saturates its visibility bound, whereas the Allen-Dynes minimum lies below its bound; the disjoint regions quantify the method dependence of the inferred couplings.

	Within this near-threshold regime and for a representative value $\mu^* = 0.13$, we use the representative parameters
	\begin{align}
		\Omega_{\log} &\approx 9.75~\text{K} \,\, (0.84~\text{meV}), \nonumber\\
		\lambda_0 &\approx 0.27, \nonumber\\
		C &\approx 2.87. \nonumber
	\end{align}
	$\Omega_{\log}$ is the representative effective spectral scale adopted within this correlated Allen-Dynes fit family and is the moment entering its prefactor. It should not be identified with the peak of $\alpha^2F(\omega)$ or described as independently determined by the numerical Migdal-Eliashberg calculation. Fixing the effective oscillator parameter at the bulk transverse-optical scale, $\Omega_B=3$ meV, the representative value $\Omega_{\log}=0.84$ meV corresponds to $\gamma\approx8.8$ meV, $\bar{\omega}_2\approx2.97$ meV, and the conditional spectral peak $\Omega_0\approx1.13$ meV. The numerical calculation subsequently holds this spectrum fixed and refits only $(\lambda_0,C)$. These values should be understood as representative parameters within the physically relevant near-threshold regime rather than as uniquely extracted quantities. The physically relevant feature is that this regime preserves a weak baseline channel together with a strong geometrically activated polar enhancement.
	
	\section{Supplementary Note 7: Scope of Mixed-Parity Gap Statements}
	At a noncentrosymmetric interface, the gap can contain mixed singlet and triplet components. Their relative magnitude depends on the momentum-dependent pairing interaction and the gaps on both helicity bands. The scalar one-band calculation determines only an integrated pairing scale and makes no prediction for the gap composition. Likewise, the angular ansatz for one dynamic channel does not determine the full static $\alpha_R(\theta)$. Orientation-resolved gap spectroscopy could constrain a momentum-dependent multiband extension, but singlet--triplet mixing alone would not uniquely identify PNR-mediated pairing.
	
	\section{Supplementary Note 8: Adiabatic Estimates and Methodological Caveats}
	At (111), the numerical Eliashberg fit gives $\lambda\approx1.72$. Taking the representative value $E_F\approx40$ meV, the conventional estimates are $\lambda\Omega_{\log}/E_F\approx0.036$ and $\lambda\Omega_0/E_F\approx0.049$ when the conditional peak is used. These small ratios motivate Migdal-Eliashberg theory as a working approximation, but do not validate it for a broad overdamped spectrum with no unique boson frequency or for the unresolved momentum structure of the interaction. The corresponding low-frequency mass-renormalization estimate is $Z^{-1}\approx1/(1+\lambda)\approx0.37$. The isotropic equations omit vertex corrections; quantifying them requires a momentum-dependent treatment beyond the present calculation.
	
	\section{Supplementary Note 9: Numerical Test of the Angular Mapping}
	The Allen-Dynes interpolation was benchmarked for conventional phonon spectra, so its extension to the present broad spectrum is not guaranteed. We test whether the angular trend can also be reproduced by numerical solutions of the reduced isotropic equations. Because the coupling parameters are refitted, this comparison is not a benchmark of Allen-Dynes accuracy at identical physical parameters.

	The logical order is important. The representative scale $\Omega_{\log}=0.84$ meV is first selected from the correlated near-threshold Allen-Dynes fit family described in Supplementary Note~6. After fixing $\Omega_B=3$ meV and thereby determining $\gamma\approx8.8$ meV, the resulting spectrum is held fixed in the numerical calculation. The Migdal-Eliashberg fit therefore determines only the conditional coupling parameters $(\lambda_0,C)$; it does not independently extract $\Omega_{\log}$.
	
		To address this, we solved the linearized isotropic Matsubara-Eliashberg equations numerically. For the assumed effective scalar intra-band coupling, the equations reduce to a 1D grid over Matsubara frequencies $\omega_n=\pi k_BT(2n+1)$. Writing $Z_n(\theta)\equiv Z(i\omega_n;\theta)$, $\Delta_n(\theta)\equiv\Delta(i\omega_n;\theta)$, and $\lambda_{n-m}(\theta)\equiv\lambda(i\omega_n-i\omega_m;\theta)$,
		\begin{equation}
			Z_n(\theta) = 1 + \frac{\pi k_BT}{\omega_n}
			\sum_m \lambda_{n-m}(\theta)\operatorname{sgn}(\omega_m),
		\end{equation}
		\begin{equation}
			\Delta_n(\theta) = \frac{\pi k_BT}{Z_n(\theta)}
		\sum_m\left[\lambda_{n-m}(\theta)-\mu_m^*\right]
		\frac{\Delta_m(\theta)}{|\omega_m|}.
	\end{equation}
	The displayed sums run over all integer Matsubara indices. Numerically, even-frequency symmetry reduces them to nonnegative frequencies: the mass-renormalization equation contains the difference of the kernels at $|\omega_n-\omega_m|$ and $\omega_n+\omega_m$, while the gap equation contains their sum and a doubled Coulomb contribution.
	Here $\mu_m^*\equiv\mu^*\Theta(\omega_c-|\omega_m|)$ with $\omega_c=30$ meV, matching the cutoff used in the numerical solver.
	Using $\alpha^2F(\omega;\theta)=\lambda(\theta)B(\omega)$ with $2\int_0^\infty B(\omega)d\omega/\omega=1$, the pairing kernel is
	\begin{equation}
		\lambda(i\nu;\theta)
		=2\int_0^\infty d\omega\,
		\frac{\omega\,\alpha^2F(\omega;\theta)}{\omega^2+\nu^2}
		=\lambda(\theta)\frac{\Omega_B^2}{\Omega_B^2+\gamma|\nu|+\nu^2}.
	\end{equation}
	
	Thus $\lambda(0;\theta)=\lambda(\theta)=\lambda_0+C\sin^2\theta$: Eq.~\eqref{eq:lambda_total} supplies the zero-frequency amplitude, and the normalized Brownian factor supplies the frequency dependence used in every matrix element. The same spectrum is applied to the effective baseline and the angular enhancement. We identify $T_c$ as the temperature at which the largest eigenvalue of the linearized gap kernel reaches unity. \textit{Numerical details.} The analytic kernel agrees with direct spectral integration, and the Matsubara sum extends to $90$ meV. Raising the latter to $180$ meV changes the 25-mK visibility coupling only from $0.39044$ to $0.39053$. The converged fit to the 23 detected transitions, with the (100) null result imposed as $T_c\leq25$ mK, gives $\lambda_0=0.390$ and $C=2.000$.
	
	As shown in Fig.~4, the numerical Migdal-Eliashberg and Allen-Dynes curves display the same quasi-linear $T_c(\theta)$ trend after their coupling parameters are refitted at the shared spectrum. This supports the robustness of the angular mapping within the reduced model. It does not constitute an independent numerical determination of the bosonic energy scale. The difference between the fitted couplings shows method dependence conditional on the adopted representative spectrum; the separation of the refitted curves is not an error estimate for Allen-Dynes at fixed coupling.
	
	\section{Supplementary Note 10: Sensitivity to $\mu^*$ and Comparison with Empirical Fits}
	
	\textbf{1. Sensitivity to the Coulomb pseudopotential $\mu^*$}
	
	In the main text, we fix $\mu^* = 0.13$ as a representative value for a strongly screened oxide interface. Because $\mu^*$ and the attractive pairing strength enter the transition-temperature calculation in a correlated way, the fitted couplings are conditional on this choice. The relevant reduced-model result is therefore not robustness of a unique parameter set, but the extended quasi-linear $T_c(\theta)$ regime generated when the nonlinear $T_c(\lambda)$ relation acts on the geometric input $\lambda(\theta)=\lambda_0+C\sin^2\theta$.
	
	To isolate the Coulomb sensitivity without introducing another covariant refit, the main text holds the shared spectrum and the boundary-saturating numerical baseline $\lambda_0=0.390$ fixed. Reducing $\mu^*$ from $0.13$ to $0.10$, $0.08$, and $0.04$ then raises the calculated (100) transition temperature from $25$ mK to approximately $40$, $55$, and $104$ mK, respectively. These are fixed-parameter sensitivity estimates, not alternative fits. Screening can also modify the polar spectrum and electron--boson coupling, so the calculation establishes only the direction of the Coulomb effect under the stated constraints, not a unique range of refitted parameters.
	
	\textbf{2. Comparison with empirical alternatives}
	
	One may ask whether simpler empirical forms can describe the same dataset without invoking the reduced Eliashberg framework. To address this, we compare the present fit against two representative alternatives:
	(i) a purely linear fit, $T_c = a\theta + b$, and
	(ii) a thresholded quadratic fit, $T_c = A \max[\sin^2\theta-s_0,0]$, where $s_0$ is a dimensionless onset threshold in $\sin^2\theta$.
	
	A linear fit to the 23 detected transitions gives $\mathrm{SSR}=0.5151$ K$^2$, but reaches this value by extrapolating to the unphysical intercept $T_c(100)=-0.750$ K; constraining the intercept to $0\leq T_c(100)\leq25$ mK raises the residual to $1.5075$ K$^2$. Thus an unthresholded line does not represent the collapse of superconductivity near (100). By contrast, the thresholded quadratic automatically satisfies the (100) bound and gives $\mathrm{SSR}=0.4790$ K$^2$, slightly below the numerical Eliashberg value $0.4893$ K$^2$. For $N=23$ and $p=2$, the corresponding residual standard errors, $s=\sqrt{\mathrm{SSR}/(N-p)}$, are $0.151$ K and $0.153$ K, respectively. It remains purely descriptive, however: it provides neither a bosonic pairing scale nor a geometric Rashba vertex or Coulomb threshold. The reduced Eliashberg framework instead unifies the threshold and quasi-linear regimes within a physically motivated pairing scenario.
	
	For transparency, Table~\ref{tab:ssr_compare} lists the raw SSR values evaluated on the same 23 detected transitions. The pristine (100) result is not entered as an artificial zero: it constrains the reduced-model fits to $T_c(100)\leq25$ mK and is satisfied automatically by the thresholded quadratic.
	
	\begin{table}[h]
		\centering
		\caption{Summed squared residuals over the 23 detected transitions for representative empirical fits and the reduced model. The linear value is the unconstrained fit, whose extrapolated (100) intercept is unphysical. The last two rows use separately fitted couplings at the shared spectrum.}
		\begingroup
		\renewcommand{\arraystretch}{1.15}
		\begin{tabular*}{0.70\textwidth}{@{\extracolsep{\fill}} l c @{}}
			\hline\hline
			Fit form & SSR (K$^2$) \\
			\hline
			Linear fit: $T_c=a\theta+b$ & 0.5151 \\
			Thresholded quadratic fit & 0.4790 \\
			Reduced model: Allen-Dynes & 0.5889 \\
			Reduced model: numerical Eliashberg & 0.4893 \\
			\hline\hline
		\end{tabular*}
		\endgroup
		\label{tab:ssr_compare}
	\end{table}
	
	The value of the reduced Eliashberg framework is therefore not that it minimizes a purely empirical residual metric, but that it reproduces the observed angular trend while simultaneously incorporating a soft polar boson, a geometrically controlled Rashba pairing vertex, and a natural threshold associated with Coulomb competition, without resorting to piecewise empirical constructions.
	
\end{document}